\documentclass[amssymb,prl,twocolumn,superscriptaddress,showpacs]{revtex4}
\usepackage{graphicx}
\usepackage{dcolumn}
\usepackage{epsfig}
\usepackage{bm}
\usepackage{color}
\usepackage[colorlinks]{hyperref}
\usepackage{verbatim}
\usepackage[sort&compress]{natbib}
\begin{document}

%\preprint{}

%\title{Nonlinear growth with the microwave intensity in radiation-induced magnetoresistance oscillations}
\title{\bf Sub-linear radiation power dependence of photo-excited resistance oscillations in two-dimensional electron
systems}

\author{Jes\'us I\~narrea}
\affiliation {Escuela Polit\'ecnica
Superior,Universidad Carlos III,Leganes,Madrid,28911,Spain}
\affiliation{Unidad Asociada al Instituto de Ciencia de Materiales, CSIC,
Cantoblanco,Madrid,28049,Spain.}
\author{R. G. Mani}
\affiliation{Department of Physics and Astronomy, Georgia State
University, Atlanta, GA 30303 U.S.A. }
\author{W. Wegscheider}
\affiliation{Universit\"{a}t Regensburg, 93053 Regensburg,
Germany}
\affiliation{Laboratorium f\"{u}r Festk\"{o}rperphysik,
ETH-Z\"{u}rich, 8093 Z\"{u}rich, Switzerland}

%\author{}
%\thanks{}
%\affiliation{}
\date{\today}
\begin{abstract}
We find that the amplitude of the $R_{xx}$ radiation-induced
magnetoresistance oscillations in GaAs/AlGaAs system grows
nonlinearly as $A \propto P^{\alpha}$ where $A$ is the amplitude
and the exponent $\alpha < 1$.
%, with $\alpha \rightarrow 1/2$ in
%the low temperature limit.
This striking result can be explained
with the radiation-driven electron orbits model, which suggests
that the amplitude of resistance oscillations depends linearly on
the radiation electric field, and therefore on the square root of
the power, $P$. We also study how this sub-linear power law varies
with lattice temperature and radiation frequency.
\end{abstract}
\pacs{73.40.-c,73.43.Qt, 73.43.-f, 73.21.-b}
%\pacs{}
%\keywords{}
\maketitle
\section{I. introduction}
Superconductivity\cite{1} and quantum Hall effects\cite{2,3} are
known to present two distinct and interesting examples of
zero-resistance states in material systems. Another interesting
example occurs in the quantum Hall two-dimensional electron system
(2DES) when the 2DES is irradiated with microwave and terahertz
band radiation in the presence of a transverse magnetic field.
Here, it becomes possible to photo-excite into zero-resistance
states in this low dimensional system.\cite{4,5}
\begin{figure}[t]
%h=here, t=top, b=bottom, p=separate figure page
\begin{center}
\leavevmode \epsfxsize=3.0 in \epsfysize=7.0 in \epsfbox
{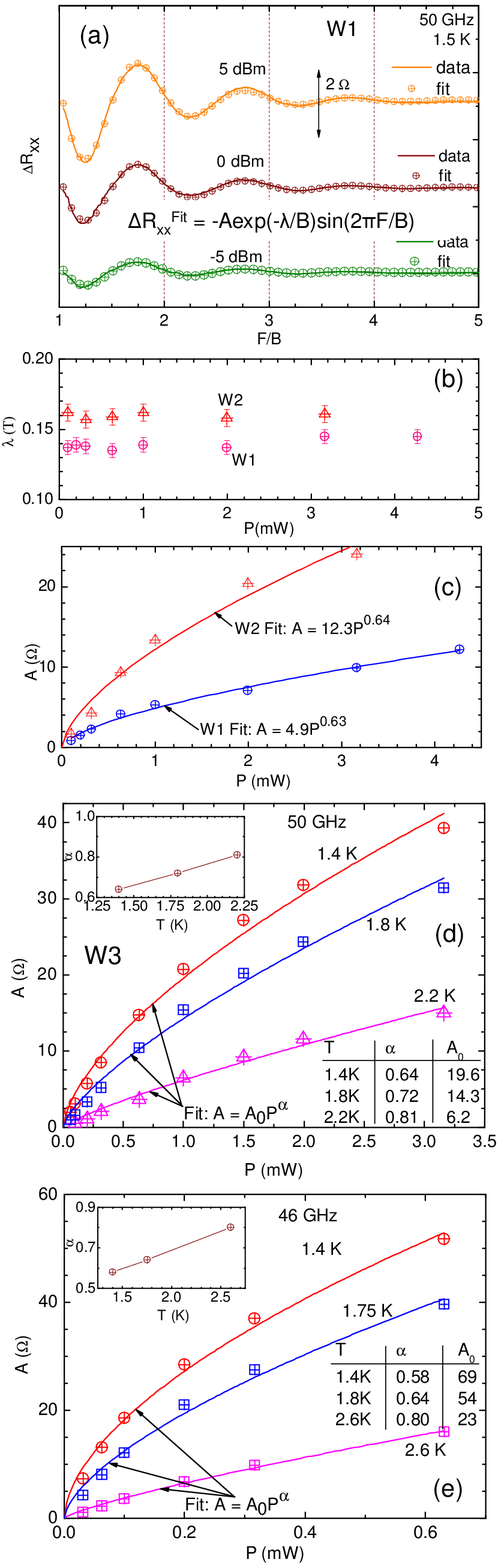}
\end{center}
%\caption{}
%\begin{comment}
\caption{(a) For W1, $\Delta R_{xx}$ is exhibited at $f=50GHz$.
Also shown are fits to an exponentially damped sinusoid. (b)
$\lambda$ is plotted vs. $P$ for W1 and W2. (c) The lineshape
amplitude, $A$, is plotted vs. $P$ for W1 and W2. Also shown are
fits, $A = A_{0} P^{\alpha}$, which suggest $\alpha = 0.63$ and
$\alpha = 0.64$ for W1 and W2, respectively.(d) At $f=50GHz$, the
amplitude, $A$, is plotted vs. $P$ for $T = 1.4K, 1.8K$ and $2.2K$
for W3. Also shown are fits to $A = A_{0} P^{\alpha}$. The
fit-extracted $\alpha$ and $A_{0}$ are presented are presented in
tabular form within the figure. The inset shows the variation of
$\alpha$ with $T$ at $50GHz$. (e) At $f=46GHz$, the amplitude,
$A$, is plotted vs. $P$ for $T = 1.4K, 1.75K$ and $2.6K$ for W3.
The inset shows the variation of $\alpha$ with $T$ at $46GHz$. }
\label{mani04fig}
%\end{comment}
\end{figure}
\begin{figure} \centering\epsfxsize=3.5in
\epsfysize=3.5in \epsffile{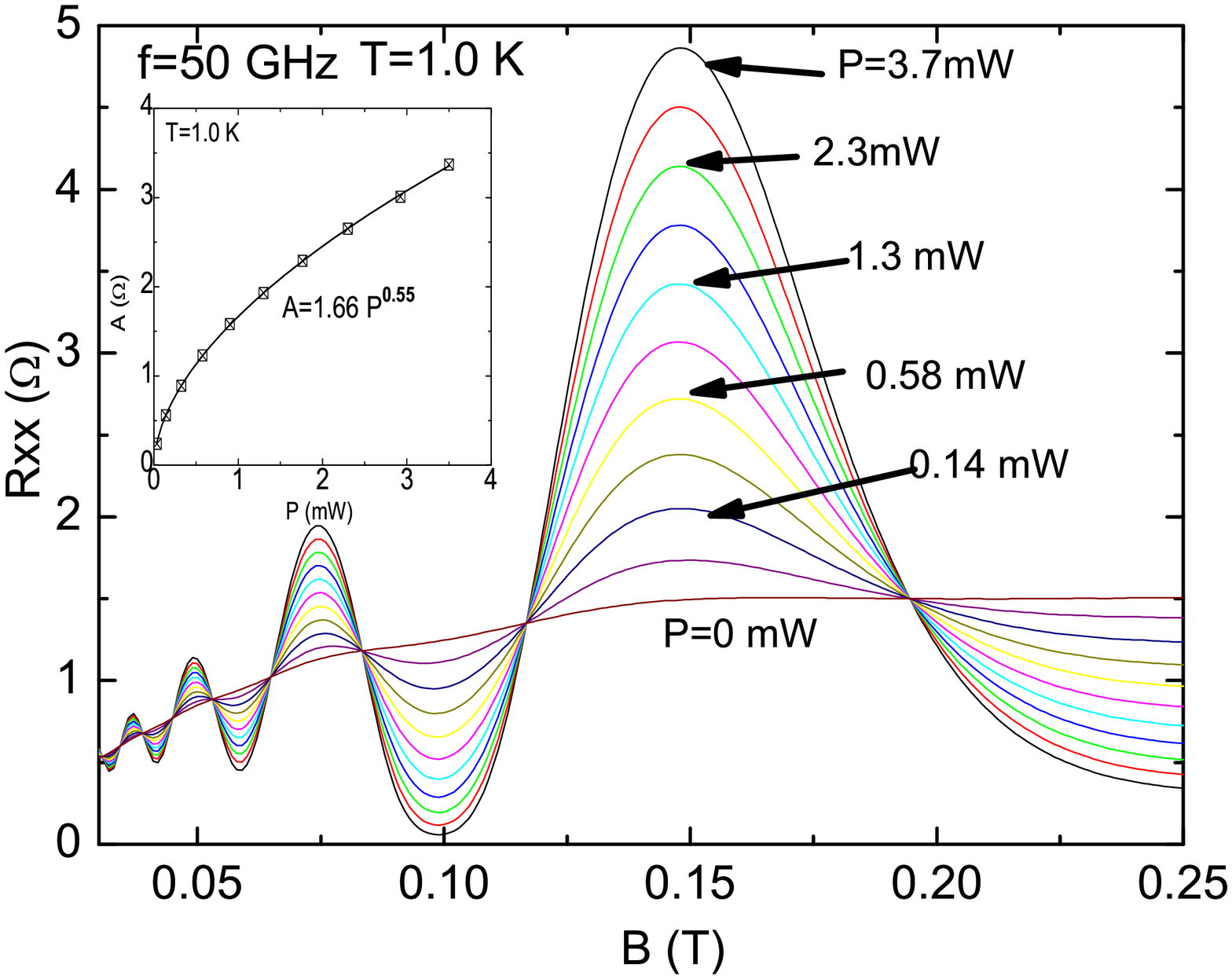} \caption{
%\begin{comment}
Calculated $R_{xx}$ vs. $B$ at $f = 50 GHz$, $T=1K$, and $0 \leq P
\leq 3.7 mW$. We obtain a progressive quenching of $R_{xx}$
oscillations with decreasing $P$, leading into the dark curve ($P
= 0 mW$). The inset shows the calculated amplitude of the main
$R_{xx}$ peak of Fig. 2 ($B=0.15T$) versus $P$. $A$ increases
sub-linearly with $P$. The line corresponds to a fit of the
calculated values with $A=1.66P^{0.55}$ which, as expected, is
very close to a square root law.}
%\end{comment}
\end{figure}
Photo-excited transport in the 2DES has become a topic of
experimental and theoretical interest since this
observation.\cite{4, 5, 6, 7, 8, 9, 10, 11, 12, 13, 14, 15, 16,
17, 18, 19, 20, 21, 22, 23, 24, 25, 26, 27,28, 29, 30, 31,32, 33,
34, 35, 36, 37, 38, 39, 40, 41, 42, 43,44} Periodic in $B^{-1}$
radiation-induced magnetoresistance oscillations, which lead into
the radiation-induced zero-resistance states, are now understood
to be a consequence of radiation-frequency ($f$) and magnetic
field ($B$) dependent, scattering at impurities \cite{ 24,25,26,
27,28} and/or a change in the distribution function\cite{6,31,38}.
And, vanishing resistance at the oscillatory minima is explained
as a feature of negative resistance instability and current domain
formation.\cite{25, 34, 42} In spite of the progress, there remain
many aspects that could be better understood from including, for
example, the growth of the oscillatory effect vs. the radiation
intensity, $P$. Here, a number of works have numerically evaluated
the radiation-induced magnetoresistance oscillations for several
$P$ and graphically presented the results,\cite{24, 32, 35} while
Dmitriev et al.,\cite{31} have suggested that the correction to
the dark dc conductivity is linear in $P$.\cite{31} A comparison
of experiment with theory, so far as the $P$-dependence is
concerned, could help to identify the importance of the
invoked-mechanisms in these theories.\cite{36}

Thus, we examine the growth of the radiation-induced
magneto-resistance oscillations with $P$. Experiment indicates
that the amplitude $A$ of the radiation-induced oscillatory
diagonal resistance ($R_{xx}$), grows nonlinearly with $P$ and can
be described by $A \propto P^{\alpha}$ where the exponent $\alpha
< 1$.
% and $\alpha \rightarrow 1/2$ in the low temperature limit.
At
the same time, according to experiment, $\alpha$ also depends on the lattice temperature
$T$ and radiation frequency $f$. Since such non-linear growth of
$A$ with $P$, $T$, and $f$ had not been predicted, we propose a
theoretical explanation. The explanation utilizes the {\it
radiation-driven electron orbit model},\cite{32} where radiation
forces the orbit center of the Landau states to move back and
forth in the direction of the radiation electric field at the
frequency of radiation, and the $R_{xx}$ oscillations reflect the
periodic motion of the electron orbits center. This theory
establishes that the amplitude ($A_{Lan}$) of the orbit center
motion, sets the amplitude $A$ of the $R_{xx}$ oscillations, and
that $A_{Lan}$ is proportional to the radiation electric field.
Since the radiation power $P$ is proportional to the square of the
radiation electric field, it follows that $R_{xx}$ and $A$ will
depend on the square root of $P$, at the lowest temperatures.
The
experiments indicate that $\alpha$ is close to
$0.5$ at lower temperatures.
%a convergence towards this value in the low
%temperature limit.

\section{II. Theoretical model}

The above mentioned theory\cite{32,33,39} was proposed to explain
the $R_{xx}$ of an irradiated 2DES at low $B$. We obtained the
exact solution of the corresponding electronic wave
function\cite{32,39,45,46,47,48}:
\begin{eqnarray}
\Psi(x,t)\propto\phi_{n}(x-X-x_{cl}(t),t)
\end{eqnarray}
%&&\times  exp
%\left[i\frac{m^{*}}{\hbar}\frac{dx_{cl}(t)}{dt}[x-x_{cl}(t)]+
%\frac{i}{\hbar}\int_{0}^{t} {\it L} dt'\right]\nonumber  \\
%&&\times\sum_{m=-\infty}^{\infty} J_{m}\left[\frac{eE_{0}}{\hbar}
%X\left(\frac{1}{w}+\frac{w}{\sqrt{(w_{c}^{2}-w^{2})^{2}+\gamma^{4}}}\right)\right]
%e^{imwt}
%\end{eqnarray}
, where $\phi_{n}$ is the solution for the
Schr\"{o}dinger equation of the unforced quantum harmonic
oscillator, $X$ is the center of the orbit for the electron
motion, $x_{cl}(t)$ is the classical solution of a forced harmonic
oscillator:
\begin{eqnarray}
x_{cl}&=&\frac{e
E_{o}}{m^{*}\sqrt{(w_{c}^{2}-w^{2})^{2}+\gamma^{4}}}\cos wt\nonumber\\
&=&A_{Lan}\cos wt
\end{eqnarray}
where $e$ is the electron charge, $\gamma$ is a
phenomenologically-introduced damping factor for the electronic
interaction with acoustic phonons, $w_{c}$ the cyclotron
frequency, $E_{0}$ the radiation electric field.  Then, the
obtained wave function is the same as the standard harmonic
oscillator where the center is displaced by $x_{cl}(t)$. Thus, the
orbit centers are not fixed, but they oscillate harmonically at
the radiation field frequency $w=2\pi f$.
Then, by this micro(MW)-driven periodic motion, electrons in their orbits (quantum
oscillators) interact with the lattice ions being damped and
emitting acoustic phonons. In the $x_{cl}$ expression, $\gamma$
represents this damping.
%Then, the amplitude, $A_{Lan}$, is given
%by:
%\begin{equation}
%$A_{Lan}=\frac{e
%E_{o}}{m^{*}\sqrt{(w_{c}^{2}-w^{2})^{2}+\gamma^{4}}}$
%\end{equation}

This $radiation-driven$ behavior affects dramatically the charged
impurity scattering and eventually the conductivity. Then, first we
calculate the impurity scattering rate $W_{N,M}$
between two $oscillating$ Landau states $\Psi_{N}$, and
$\Psi_{M}$\cite{32,47,49,50}.
%Now we introduce the scattering suffered by the electrons due to
%charged impurities randomly distributed in the sample.
%Electrons suffer scattering due to charged impurities that are
%randomly distributed in the sample\cite{47}. If the scattering is weak, we
%can apply time dependent first order perturbation theory.
%To
%proceed we calculate the impurity scattering rate $
%W_{N,M}$ from an initial state $\Psi_{N}$, to a final state
%$\Psi_{M}$\cite{32,49}.
%\begin{equation}
%W_{N,M}=\lim_{\alpha\rightarrow 0} \frac{d}{d t} \left|
% \frac{1}{i \hbar} \int_{-\infty}^{t^{'}}<\Psi_{M}(x,t) |V_{s}|\Psi_{N}(x,t)>e^{\alpha t}d t\right|^{2}
%\end{equation}
%where $V_{s}$ is the scattering potential for charged
%impurities\cite{47,50}.
%Firstly we calculate the electron-charged impurity scattering rate
%$1/\tau$, being $\tau$ the scattering time.
Next we find the average effective distance advanced by the electron
in every scattering jump: $\Delta X^{MW}\propto A_{Lan}\cos w\tau$\cite{32,39,47},
where $\tau=1/W_{N,M}$ is
the scattering time. Finally the longitudinal conductivity
$\sigma_{xx}$ is given by:
\begin{equation}
\sigma_{xx}\propto \int dE \frac{\Delta X^{MW}}{\tau}
\end{equation}
being $E$
the energy.
%\end{equation}
%being $f_{i}$ and $f_{f}$ the corresponding distribution functions
%for the initial and final Landau states respectively and $E$ energy.
%$R_{xx}$ is obtained from $\sigma_{xx}$ by the usual tensorial
%relationships\cite{32,39}.
%\begin{figure}
%\centering\epsfxsize=3.5in \epsfysize=2.7in
%\epsffile{manipowerfig2.eps} \caption{Calculated amplitude of the
%main $R_{xx}$ peak of Fig. 1 ($B=0.15T$) versus $P$. The radiation
%frequency and $T$ are the same as in Fig. 1. We obtain a clear
%sublinear relation between $A$ and $P$. The single line corresponds
%to a fit of the calculated values obtaining $A=1.66P^{0.55}$ which,
%as expected, it is very close to a square root law.}
%\end{figure}
To obtain $R_{xx}$ we use
the relation
$R_{xx}=\frac{\sigma_{xx}}{\sigma_{xx}^{2}+\sigma_{xy}^{2}}
\simeq\frac{\sigma_{xx}}{\sigma_{xy}^{2}}$, where
$\sigma_{xy}\simeq\frac{n_{i}e}{B}$ and $\sigma_{xx}\ll\sigma_{xy}$.
Therefore,
\begin{equation}
R_{xx}\propto \frac{e
E_{o}}{m^{*}\sqrt{(w_{c}^{2}-w^{2})^{2}+\gamma^{4}}}\cos wt=A_{Lan} \cos w\tau
\end{equation}
and the
amplitude of resistance oscillations depends linearly
on the radiation electric field.

$P$ can be related with
$R_{xx}$ through the well-known formula that gives radiation
intensity $I$ (power divided by surface) in terms of the radiation
electric field $E_{0}$: $I=\frac{1}{2}c\epsilon_{0}E_{0}^{2}$, where
$c$ is the speed of light in vacuum and  $\epsilon_{0}$ is the
permittivity in vacuum. If we want to express only the power in
terms of the radiation electric field we have to take into account
the sample surface. In the particular case of GaAs if $S$ is the
sample surface, we can readily obtain:
\begin{equation}
 P = \frac{1}{2} c_{GaAs} \epsilon_{GaAs}\epsilon_{0}E_{0}^{2} S
\end{equation}
where $c_{GaAs}$ is the speed of light in GaAs and $\epsilon_{GaAs}$
is the GaAs dielectric constant.  Accordingly, $E_{0}\propto
\sqrt{P}$. Then, substituting in the expression of $A_{Lan}$, we
obtain that $R_{xx}$ varies with $P$ following an square root law:
\begin{equation}
R_{xx}\propto \frac{e\sqrt{P}
}{m^{*}\sqrt{(w_{c}^{2}-w^{2})^{2}+\gamma^{4}}}\cos w \tau
\end{equation}

Thus, the $R_{xx}$ response will grow as the square root of $P$,
i.e., $A\propto P^{0.5}$. From the expression of $R_{xx}$ we can
study the influence of $w$ and $T$ on the sublinear relation. In
the case of $w$, the calculation is straightforward (see
expressions of $R_{xx}$ or $A_{Lan}$). In the case of  $T$, we
introduce a microscopic model, which allow us to obtain the
damping parameter $\gamma$ and its dependence on $T$
\cite{32,33,50}. According to the model, $\gamma$ is proportional
to the scattering rate of electron-acoustic phonon interaction,
being eventually linear with $T$\cite{50}.
%\begin{equation}
%$\gamma \propto \frac{1}{\tau_{ac}}$.
%\end{equation}
%$\frac{1}{\tau_{ac}}$ can be calculated using
%the Fermi's Golden rule obtaining an expression that implies a linear
%relation with $T$\cite{51,52}.
%\begin{equation}
%\frac{1}{\tau_{ac}}=\frac{m^{*}\Xi_{ac}^{2}k_{B}T}{\hbar^{3}\rho
%u_{l}^{2}<z>}
%\end{equation}
%where $\Xi_{ac}$ is the acoustic deformation potential, $\rho$ the
%mass density, $u_{l}$ the sound velocity and $<z>$ is the effective
%layer thickness.
% Finally, $\gamma$ depends linearly on $T$ through the probability rate
%$\frac{1}{\tau_{ac}}$, thus, $\gamma \propto T$.
\section{Experimental set up and results}

Low frequency lock-in based electrical measurements were carried
out at $T \leq 1.5 K$ with the samples immersed in pumped
liquid-helium and mounted near the open end of a microwave
waveguide.\cite{4,13}  The W (Wegscheider)-GaAs/AlGaAs single
heterostructures were nominally characterized by an electron
density, $n = 2.4 \times 10^{11} cm^{-2}$ and a mobility of $\mu =
10^{7} cm^{2}/Vs$. Results are reported here for measurements on a
set of three W-specimens.

%Figure 1(a) exhibits the data for a $0.4 mm$ wide Hall bar (W1)
%fabricated from the W-material. The figure shows the $R_{xx}=
%V_{xx}/I$ measured both in the dark (w/o radiation) and under $50
%GHz$ photo-excitation (w/ radiation).\cite{4,23} In Fig. 1(b),
%$R_{xx}$ on the right ordinate has been plotted vs. $B^{-1}$ in
%order to exhibit the $B^{-1}$ periodicity of the oscillations. A
%half-cycle analysis was carried out as in ref.\cite{8}. The
%analysis indicates an intercept $k_{0} = 0.22$, see Fig. 1(b),
%which confirms a "1/4-cycle phase shift".\cite{8} In addition, the
%fit-slope indicates that $ m^{*}/m  = 0.0657$, slightly lower than
%the standard value, $m^{*}/m = 0.067$, for the GaAs/AlGaAs
%system.\cite{8}

To examine the growth of the radiation-induced oscillations with
$P$, Fig. 1 presents the $\Delta R_{xx}$ of specimen W1 for
several $P$ at $50GHz$. Also shown in the figure are data-fits to
$\Delta R_{xx}^{fit} = -A exp(-\lambda/B)sin(2 \pi F/B)$. Here, a
slowly varying background, approximately equaling the dark trace,
was removed from the photo-excited $R_{xx}$ data to realize
$\Delta R_{xx}$.\cite{29} Although this fit function includes
three parameters, $A$, $\lambda$, and $F$, the oscillation period
in $B^{-1}$ is independent of the radiation-intensity, and thus,
$F$ is a constant. Further, the damping constant,
$\lambda$,\cite{9} turns out to be insensitive to $P$, see Fig.
1(b). Thus, the main free parameter in the fit-function is the
amplitude, $A$, of the oscillations. In Fig. 1(c), we exhibit the
fit extracted $A$ vs. $P$ for W1 and W2. The figure shows a
sub-linear growth of $A$ with $P$. Also shown are power law fits,
$A = A_{0}P^{\alpha}$. Here, $\alpha = 0.63$ and $\alpha = 0.64$
for W1 and W2, respectively, at $1.5K$ and $50GHz$. $A_{0}$ varies
between W1 and W2 because the effective intensity attenuation
factor varies between the two experiments.

For a third W-specimen labeled W3, Fig. 1(d) and 1(e) report the
influence of the temperature on the growth of $A$ vs. $P$, where
$A$ is extracted, as before, from line-shape fits of the
oscillatory data to $\Delta R_{xx}^{fit} = -A exp(-\lambda/B)sin(2
\pi F/B)$. Here, Fig. 1(d) and 1(e) indicate that, as expected, at
a constant $P$, $A$ grows with decreasing $T$ both at $f=50 GHz$
[Fig. 1(d)] and $f=46GHz$ [Fig 1(e)]. Further, the figures show
that the $A$ vs. $P$ curves exhibit greater curvature at lower
temperatures. The $A$ vs. $P$ have been fit once again to $A =
A_{0}P^{\alpha}$; the fit-extracted $\alpha$ and $A_{0}$ have been
summarized in tabular form within Fig. 1(d) and Fig. 1(e). These
fit-extracted $\alpha$ have also been plotted vs. $T$ in the inset
to these figures. These insets suggest that $\alpha$ decreases
with decreasing temperatures, consistent with the observed
increased non-linearity at lower temperatures. Note that, at
$f=50GHz$, all three W-specimens, exhibit the same $\alpha$,
within uncertainties, at the lowest pumped $^{4}$He temperatures.
In addition, a comparison of the $\alpha$ reported in Fig. 1(d)
and Fig. 1(e) also suggests that reducing the microwave frequency
$f$ at a fixed $T$ tends to reduce the $\alpha$, i.e., increase
the non-linearity. Thus, the experimental results presented here
suggest a nonlinear power law in a regime characterized by modest
excitation. This peculiar behavior can be theoretically explained
and the experimental results recovered by means of the
radiation-driven electron orbits model. Next we present the
calculated results that seem to reasonably agree with the present
experiments.

\section{Calculated results}

Fig. 2 presents the calculated $R_{xx}$ as a function of  $B$ over
$0 \le P \le 3.7 mW$ at $f = 50GHz$ and $T=1K$.
As
in experiments, we obtain multiple oscillations for all $P$
except, as expected, for $P=0$. It is clear the progressive
collapse of $R_{xx}$ oscillations as $P$ decreases. The
explanation is straightforward according to our theory and it is
given by the expression of $R_{xx}$. Thus, for a decreasing
(increasing) $P$ the $R_{xx}$ response is progressively smaller
(larger).
The key issue here, and the challenge for the different
available theories, is to deduce the  rate of increase (decrease)
of $R_{xx}$ oscillations for increasing (decreasing) $P$. This is
shown in the inset of Fig. 2, where we present the calculated
amplitude versus $P$.  As in experiments, we obtain a clear sublinear growth
of $A$ with $P$: $A\propto P^{\alpha}$. We would expect an exponent $\alpha$ around
$0.5$ since the model establishes an square root dependence. Thus,
 we have performed a fit of the calculated
values, obtaining $A=1.66P^{0.55}$, very
close to a square root law.

\begin{figure}
\centering\epsfxsize=3.5in \epsfysize=5.2in
\epsffile{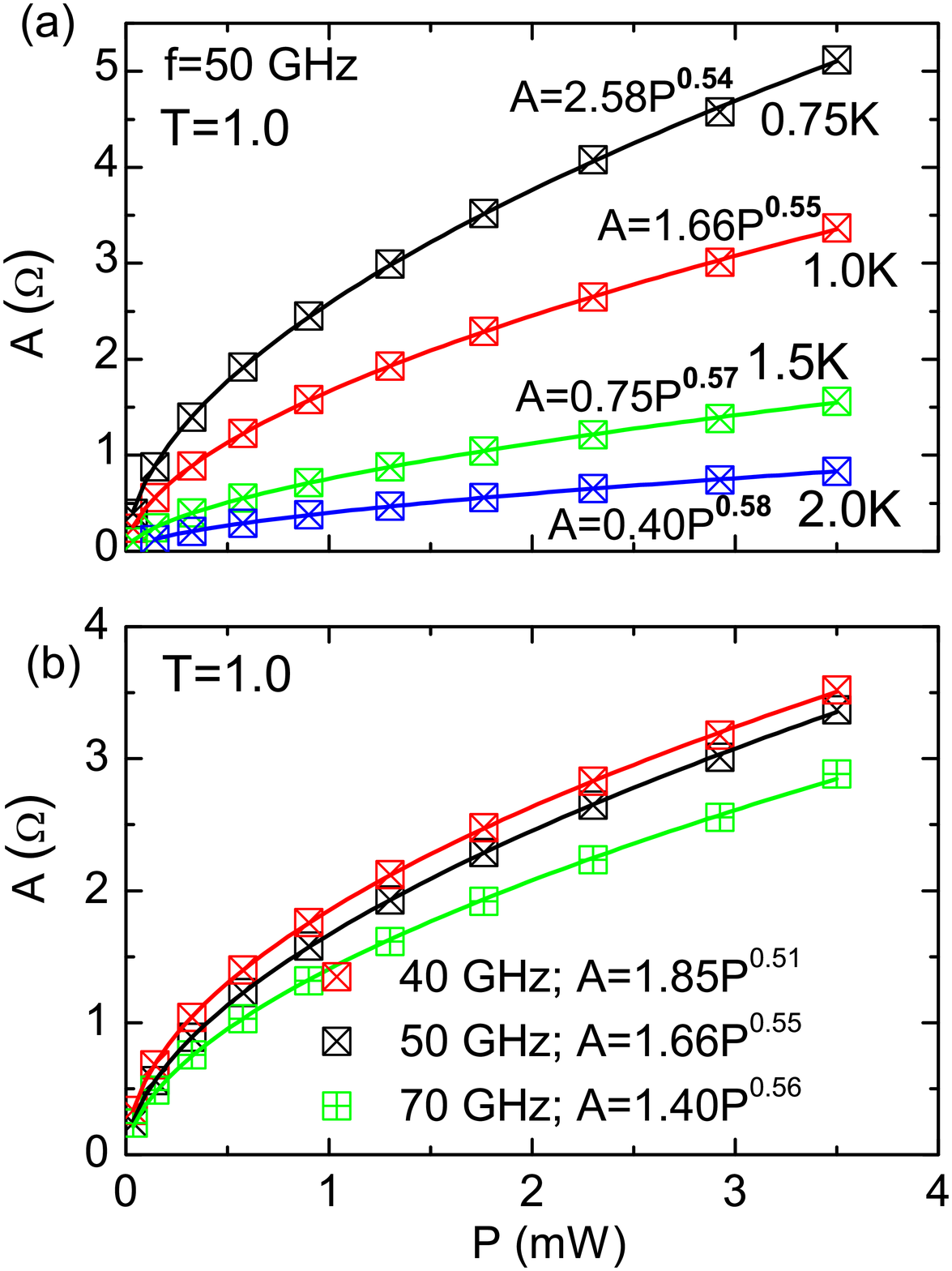} \caption{
%\begin{comment}
(a) Calculated amplitude $A$ of the main $R_{xx}$ peak versus $P$
for $f = 50GHz$ and the $T$ values: $T=0.75K$, $1.0K$, $1.5 K$ and
$2.0K$. Note that the dependence between $A$ and $P$ is sub-linear
for all $T$. Also shown are the fits performed for each $T$.
Observe that for increasing $T$, the exponent of the fit increases
as the pre-factor decreases.(b) Calculated $A$ versus $P$ for
radiation frequencies, $40GHz$, $50GHz$ and $70Ghz$ and $T=1K$. We
observe that $A$ decreases as $f$ gets larger. A fit of the $A$
vs. $P$ curves indicates that the exponent increases with
increasing frequency, whereas the pre-factor decreases.}
%\end{comment}
\end{figure}
The dependence of the sublinear law on $T$ and frequency
 is presented in Fig. 3.
In Fig. 3a we represent the calculated amplitude $A$ of the main
$R_{xx}$ peak versus $P$ for a frequency of $50GHz$ and
$T$ values: $T=0.75K$, $1.0K$, $1.5 K$ and $2.0K$. We obtain
a qualitatively similar behavior  as in experiments. Thus, we observe that
for a constant $P$, $A$ increases (decreases) with
decreasing (increasing) $T$. The physical explanation is a follows.
If one increases $T$, the interaction between
electrons and lattice ions also increases giving rise to a more intense damping.
This implies progressively smaller amplitudes and $R_{xx}$
oscillations tend to vanish. On the other hand, it is very clear
that for all $T$ used in the calculation, the dependence between
$A$ and $P$ is sub-linear and that, at least qualitatively,
the curves show a bigger curvature for decreasing $T$.
To check this curvature, we have made a fit to the calculated
values for each $T$  observing that the exponent of the
fits slightly increases for increasing $T$. In all cases, the
exponent is always close to 0.5. This outcome is expected since,
strictly speaking, our model does not predict any variation of the
exponent with $T$; our theoretical framework imposes a square root
relation. Yet, experimental results show a much faster increase of
the exponent with increasing $T$ regarding the calculated ones. The latter
show a much slower variation with $T$, although in the same direction.
At present, this model can not explain this quantitative discrepancy of
the variation of $\alpha$ with $T$. This discrepancy could be the result of
an statistical effect of the  fit or the reflex of a real physical process.
In the latter case, this would imply an extension of the current theory
that should start from the square root law obtained here. We consider
that the  square root dependence is a solid result reflecting a reasonably
good agreement between theory and experiments a low temperatures.
%Therefore, the agreement between calculated and
%experimental results in terms of $\alpha$ variation with $T$ is mainly qualitative.

%In the case of the prefactor, our theory
%predicts an increase with decreasing $T$. This behavior, which is similar
%as the one obtained in experiments,  can be observed in  the Fig. 3a.

%Yet, all
%calculated values at each $T$ have been obtained applying the same
%theoretical framework, that imposes a square root relation
%($\alpha  = 0.5$).
%This apparent contradiction indicates that, as $A$ is getting
%smaller  for increasing $T$, the curve tends slightly to behave as
%a straight line, effectively losing the non-linearity.
%This effect
%would be similar to the behavior of a square root function
%$y=\sqrt{x}$ that tends to a straight line for
%smaller $x$ values. In our particular case this effect is
%caused by the progressively smaller pre-factor  in the
%fits.
%, which reflects the decrease in $A$  for increasing $T$.
A similar behavior is presented in Fig. 3b where we show the
calculated values of $A$ versus $P$ for different
frequencies.
We observe that, as in experiments, $A$ decreases as the the frequency increases.
 The explanation comes from the expression of
$R_{xx}$ where $w$ shows up in the denominator.
Then, for a
constant $P$, a larger $w$ gives rise to smaller amplitudes.
We have fit these curves and, as expected, the exponent of the fit
slightly increases for increasing frequency whereas the pre-factor decreases.
Yet, experiments show a much faster increase.
This discrepancy is similar to the case of the $T$-dependence.
Thus, the explanation given in Fig. 3a, applies identically here
too.

All calculated results presented in Figs. 2 and 3, have been
obtained directly through equation (6).

\section{Conclusions}
In summary, experimental results indicate a nonlinear growth in
the amplitude $A$ of radiation-induced magneto-resistance
oscillations with $P$. These results can be explained with the
radiation-driven electron orbits model. According to this model,
the amplitude of the radiation-induced oscillations should be
proportional to the square root of radiation power, which implies
a sub-linear relation.  We have also studied how this
sub-linear law varies with lattice temperature and radiation
frequency obtaining only a qualitative good agreement between theory and experiments.
The obtained quantitative discrepancies can not be explained by
our theory. The origin of them could be an  statistical effect or
a real physical effect. In the latter case, this would imply an extension of
the present theoretical model, always starting from the square root law here obtained.

R.G.M. is supported by the Army Research Office under
W911NF-07-01-0158, and the Department of Energy under
DOE-DE-SC-0001762. J.I. is supported by the MCYT (Spain) under
grant: MAT2008-02626/NAN.

\end{document}